# Line of critical states for Lennard-Jones fluids

David M. Heyes* and Leslie V. Woodcock**

* Department of Physics, Royal Holloway College, University of London, Egham, Surrey, TW20 OEX, UK.

** Manchester Institute of Biotechnology, University of Manchester, 131 Princess Street, Manchester M1 7DN, UK.

## ABSTRACT

**A method of accurately obtaining the critical temperature ($T_c$) from within a mesophase of the supercritical isotherms is described. We report high-precision thermodynamic pressures from MD simulations for more than 2000 state points along 7 near-critical isotherms, for system sizes N= 4096 and 10976 of a Lennard-Jones fluid. We obtain $k_BT_c/\varepsilon$ =1.3365$\pm$0.0005 and critical pressure $p_c\sigma^3/\varepsilon$ = 0.1405$\pm$0.0002 which remains constant between two coexisting densities $\rho_c$(gas)$\rho r_o^3$= 0.370$\pm$0.01 and $\rho_c$(liquid) $r_o^3$ = 0.527$\pm$0.01 determined by supercritical percolation transition loci. Direct evaluation of Gibbs chemical potential along the critical isotherm reaffirms "a line of critical states" and non-existence of van der Waals "critical point". An analysis of the distribution of clusters from MD simulations along a supercritical isotherm (T*=1.5) gives new insight into the supercritical boundaries of gas and liquid phases, and the nature of the supercritical mesophase.**

e-mail
* david.heyes@rhul.ac.uk
** les.woodcock@manchester.ac.uk

## INTRODUCTION

The existence of a 'critical point' terminating the vapor-liquid coexistence curves in the density-temperature plane was postulated by van der Waals 130 years ago [1]. Despite extensive research by experimental, theoretical, and latterly simulation studies, the hypothesis has remained unsubstantiated. In all the intervening years the 'critical point' has not been directly measured, or directly computed; it has only been obtained by extrapolation of the coexisting vapor and liquid densities, parameterized assuming its existence with the law of rectilinear diameters [2].

Investigations into percolation transitions, and simulation studies on hard-sphere and square-well fluid equations of state [3], however, shows a coexistence 'line of critical states', between two intersecting percolation transition loci, rather than a singular point. A re-examination of the density surface of real fluids, such as argon [4,5] and water [4], has also shown that, at the critical temperature, the experimental p-V-T data on real liquids is consistent with the absence of any van der Waals 'critical point'. The present description based upon 'a line of critical states, is consistent both with Gibbs phase rule and with our knowledge of percolation transitions [5] that dates back 25 years [6-8]. van der Waals concept of a continuous cubic equation-of-state and its 'critical point' is inconsistent with Gibbs phase rule and supercritical phase transitions.

Our new self-consistent interpretation is that, at $T_c$, there is a 2-phase coexistence line between the densities of two percolation transitions. All state-points along the critical isotherm correspond to a different density, and since $(Vdp)_T = d\mu = 0$ there is a connecting line of states at $T_c,p_c$ of constant Gibbs chemical potential ($\mu$). Thus, the $2^{nd}$-order percolation transitions on intersection in the p-T plane, become a $1^{st}$-order phase transition, in compliance with Gibbs phase rule. The first compelling experimental evidence for such a flat top to the coexistence plot in the density-temperature plan for simple fluids was obtained 60 years ago by Weinberger and Schnieder from experiments on xenon [9] .

Percolation transitions in systems of hard-spheres and square wells are well-defined by a characteristic distance, which unambiguously divides configuration space into "sites" and "holes". The densities at which clusters of occupied sites or unoccupied holes, first span the whole system are referred to as percolation transitions, and denoted by PB and PA, referring to "bonded cluster" and "available volume" respectively. For real fluids, and model fluids such as the Lennard-Jones system, however, there is no such well-defined distance to define PA and PB. Yet, all the experimental evidence suggests that these percolation loci exist as higher-order discontinuities on the Gibbs density surfaces in the supercritical region, and at a sufficient low temperature, the pressures become equal to trigger the $1^{st}$ order gas-liquid condensation at $T_c$.

The available volume ($V_a$) percolation transition (PA) occurs at the density ($\rho_{pa}$) at which the volume accessible to any single mobile atom, in static equilibrium configuration of all the other atoms, percolates the whole system. It is related to the excess Gibbs chemical potential relative to an ideal gas at the same temperature and volume ($\mu_e$) by the equation

$$\mu_e = -k_B T \log_e (V_a / V) \qquad (1)$$

The available volume percolation transition (PA) can be seen as the point whence the "holes" that give rise to the chemical potential, first coalesce to become just one large hole that spans the system. The bonded cluster percolation transition (PB) occurs at the density at which the molecules themselves, or occupied sites, first give rise to a cluster that can span the system. There appears to be a relationship between the respective cluster distributions of holes and sites. The available volume percolation transition density PA is also close to the lowest density for which a single cluster of sites, spans the system, i.e. liquid-like configuration.

At percolation transitions, thermodynamic state functions can change form due to sudden changes in state-dependence of density and/or energy fluctuations. This gives rise to weak $2^{nd}$-order thermodynamic phase transitions, in which there are discontinuities in second derivatives of chemical potential with respect to temperature or pressure, notably: isothermal compressibility $(d_2\mu/dp^2)_T$, heat capacity $(d_2\mu/dT^2)_p$ and thermal expansivity $(d_2\mu/dpdT)$ all of which undergo some degree of change at a percolation transitions. Here we show that it is the intersection of two such weak second-order percolation transition that leads to first-order critical condensation and a line of critical states rather than any 'critical point'.

## CRITICAL TEMPERATURE

More than 25 years ago, studies of percolation transitions in the Lennard-Jones fluid [6] revealed evidence of a supercritical percolation line which terminated at the density on the liquid side of the gas-liquid coexistence line. This line can now be identified as the available volume percolation transition (PA) [4]. Here we re-examine the thermodynamic state functions of the Lennard-Jones fluid in the vicinity of the critical temperature to an unprecedented level of accuracy. We report new MD computations of the Lennard-Jones fluid, the scalable energy ($\varepsilon$) and distance corresponding to the diameter of a hard sphere reference fluid where $r_0$ is in dimension of the distance of zero force at $\varepsilon = -1$. Simulations were carried out using explicitly the Lennard-Jones potential for pair separations up to $2.5\sigma$, and then tapered to zero at $2.65\sigma$ using the Morris-Song scaling construction [10].

$$\phi(r) = \varepsilon\,[(r/r_o)^{-12} - 2(r/r_o)^{-6})] \qquad (2)$$

[Note that $r_o = 2^{1/6}\sigma$ : we here define reduced density $\rho = Nr_o^3/V$ for direct comparisons with hard-sphere reference model properties, but we have retained dimensions $\varepsilon/\sigma^3$ for all pressures for easy comparison with previous L-J literature data]

**Figure 1** shows pressure-density isotherms for various isotherms listed in the figure caption. The isotherms in Fig. 1 are obtained from new simulation data p-V-T data from N-V-T molecular dynamics simulations for for systems of N=4999 and N=10976 L-J atoms. Every point in Figure 1 is an average over 2000 MD time-steps of an equilibrated independent simulation. For every isotherm, the pressure at 300 state density points was evaluated.

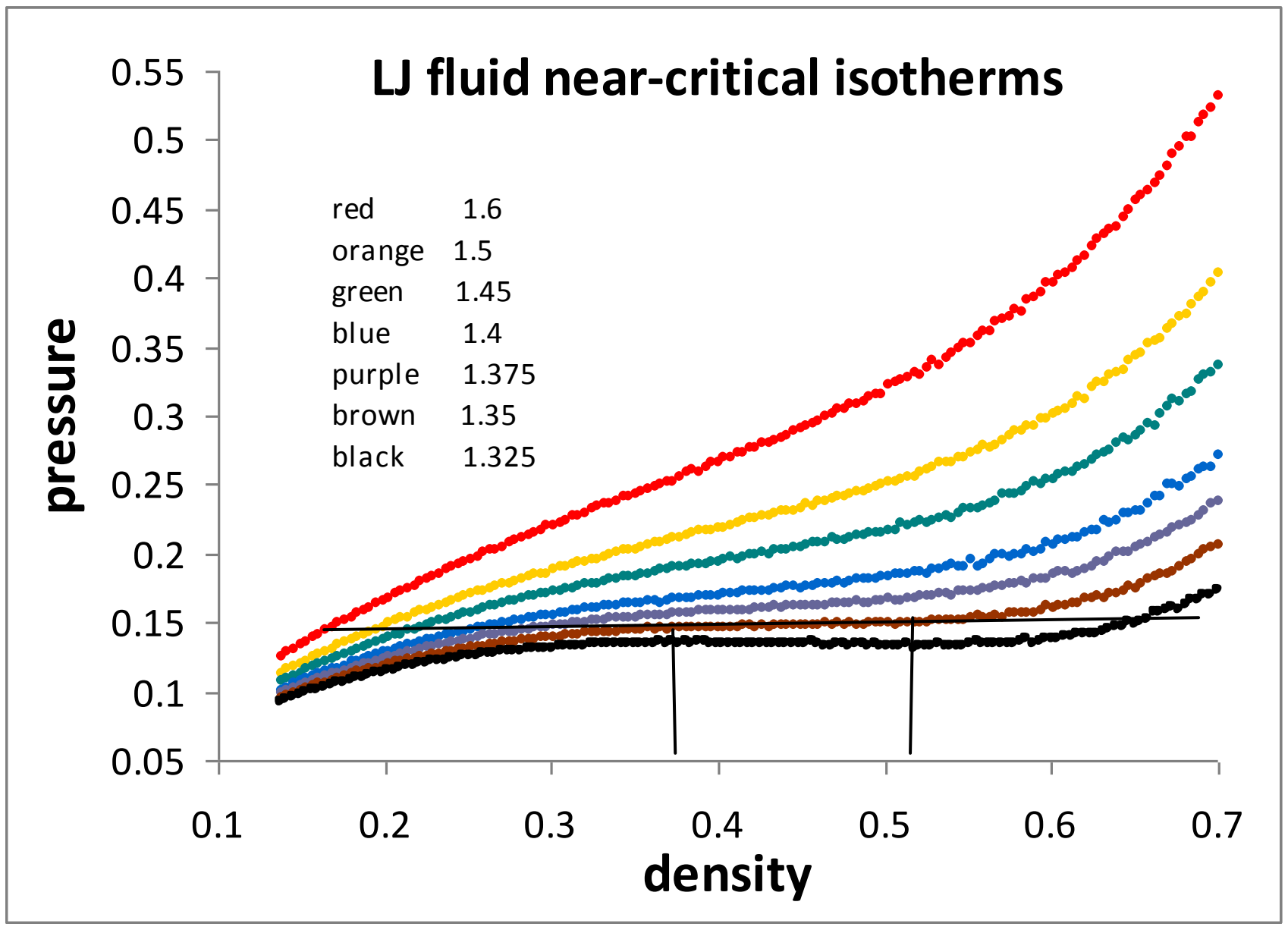

**Figure 1:** Pressure ($p\sigma^3/\varepsilon$) against density ($Nr_o^3/V$) for a series of isotherms ($k_BT/\varepsilon$) of the Lennard-Jones fluid in the vicinity of the critical temperature ($T^*_c$): all the isotherms are linear in a central region as illustrated at T*=1.35; the vertical lines indicate the densities of the percolation transitions PB and PA for the near critical isotherm (T*=1.35)

The two percolation transitions can be observed on all the supercritical isotherms as a discontinuity in the slope of the $p(\rho)$ isotherms. All the isotherms show three distinctly different regions of behavior for the state function $p(\rho)$. For $\rho < \rho_{pb}$ (gas region) the compressibility $(dp/d\rho)_T$ decreases with density, for $\rho_{pb} < \rho < \rho_{pa}$ (meso region) $(dp/d\rho)_T$ is constant, and for $r > \rho_{pa}$ $(dp/d\rho)_T$ increases with density; the slope of the meso region becomes zero at the critical temperature.

This density range within the supercritical meso phase is focused upon in **Figure 2**. It is evident from the two near critical isotherms $T_c$ = 1.35 very slightly supercritical, and $T_c$ = 1.325, very slightly subcritical, that the critical isotherm is a horizontal straight line in this region. Moreover, for the near critical isotherms (T*=1.35 and T*=1.325) both system sizes (N=4000 and N=10976) were investigated and there is no significant difference between the average pressures. Thus, in the critical region, for homogeneous single phase states, there is no unusual number dependence. The linear regression lines are shown for the intermediate density range 0.4 to 0.5, i.e. well-within the two percolation transitions for all the isotherms shown.

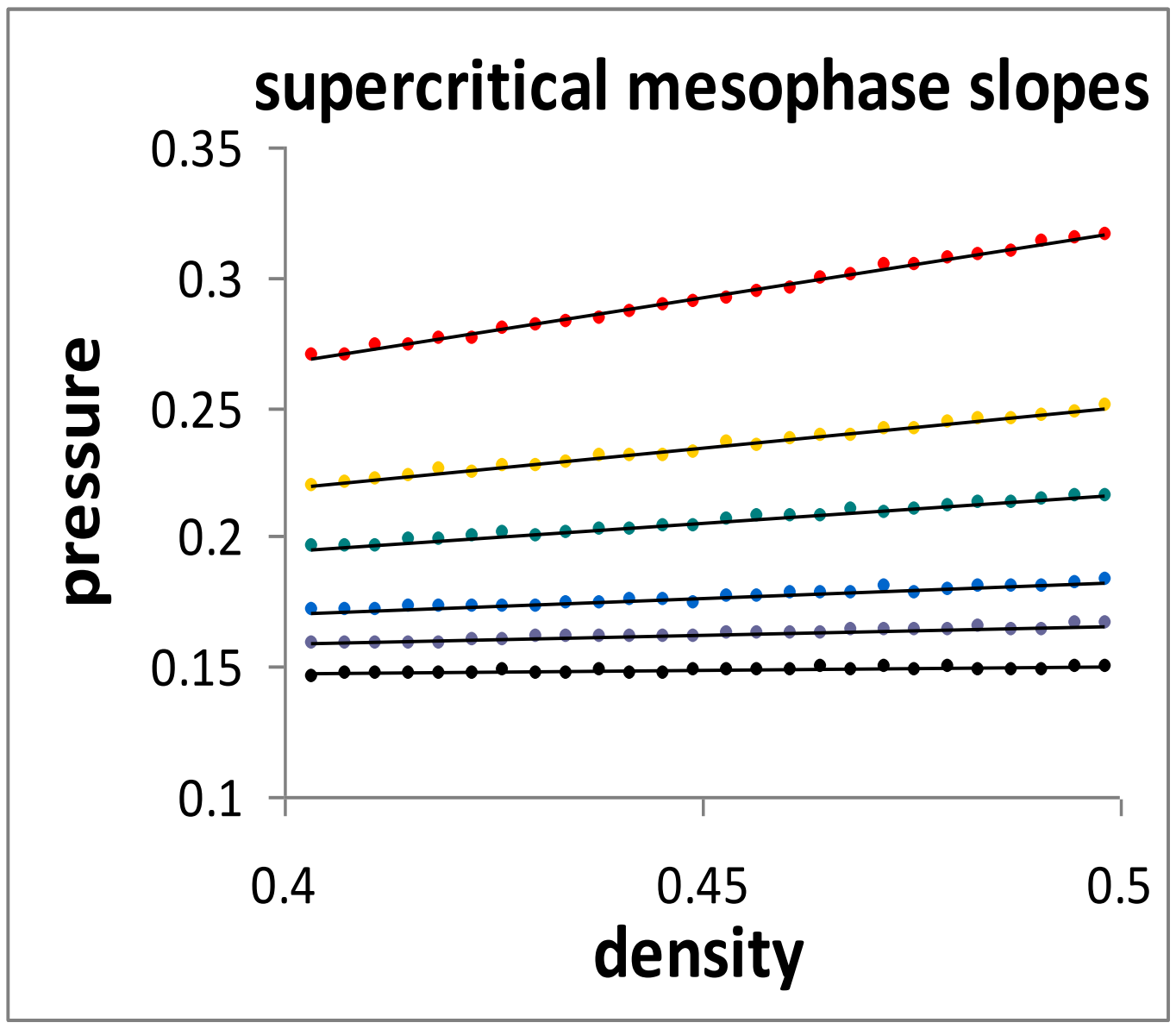

**Figure 2**: Expanded plot of the 6 supercritical isotherms shown in Figure 1 for a region within the supercritical mesophase; the plots show that in this region, the slope is linear.

The slopes of the pressures $(dp/d\rho)_T$ can be termed the 'rigidity', and have been obtained from the slopes of the isotherms in the supercritical mesophase. The rigidities in the meso-region decrease linearly with temperature, and become zero at the critical temperature. Then again using the EXCEL linear trendline (**Figure 3**), when the slope is equal to zero, a value of the reduced critical temperature $T_c^*$ is obtained.

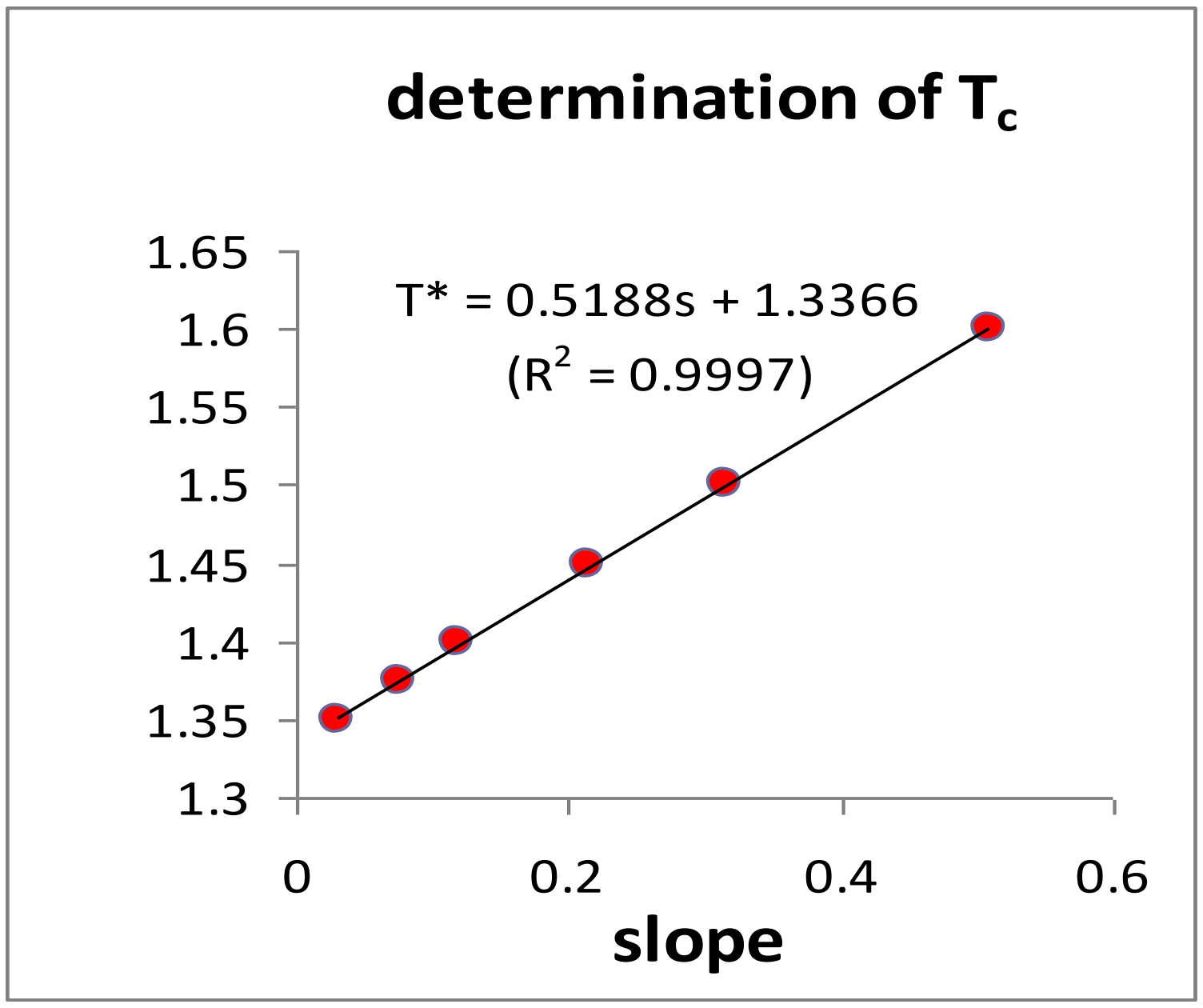

**Figure 3**: Determination of the critical temperature of the Lennard-Jones fluid.

## CRITICAL PRESSURE

Following the determination of the critical temperature ($T_c$*= 1.3365±0.0005) we can now determine an accurate critical pressure. MD computations have been carried out along the critical isotherm; high-precision values of the thermodynamic state functions of pressure and chemical potential have been computed directly for 85 density state points. The resultant critical pressure shows no van der Waals like cubic node; it is very clearly a horizontal straight line between PB and PA (**Figure 4**). The critical pressure is calculated from the average pressure of 20 state points lying between the densities 0.4 and 0.5 on the critical isotherm for a system of N=10976 L-J atoms with periodic boundaries. A precise value of $p\sigma^3/\varepsilon$= 0.1405 is obtained with an uncertainty ± 0.0002 according to the standard deviation. The smaller system of N = 4096 gives the same value to within the statistical uncertainty, again showing no significant number dependence to vitiate our conclusions.

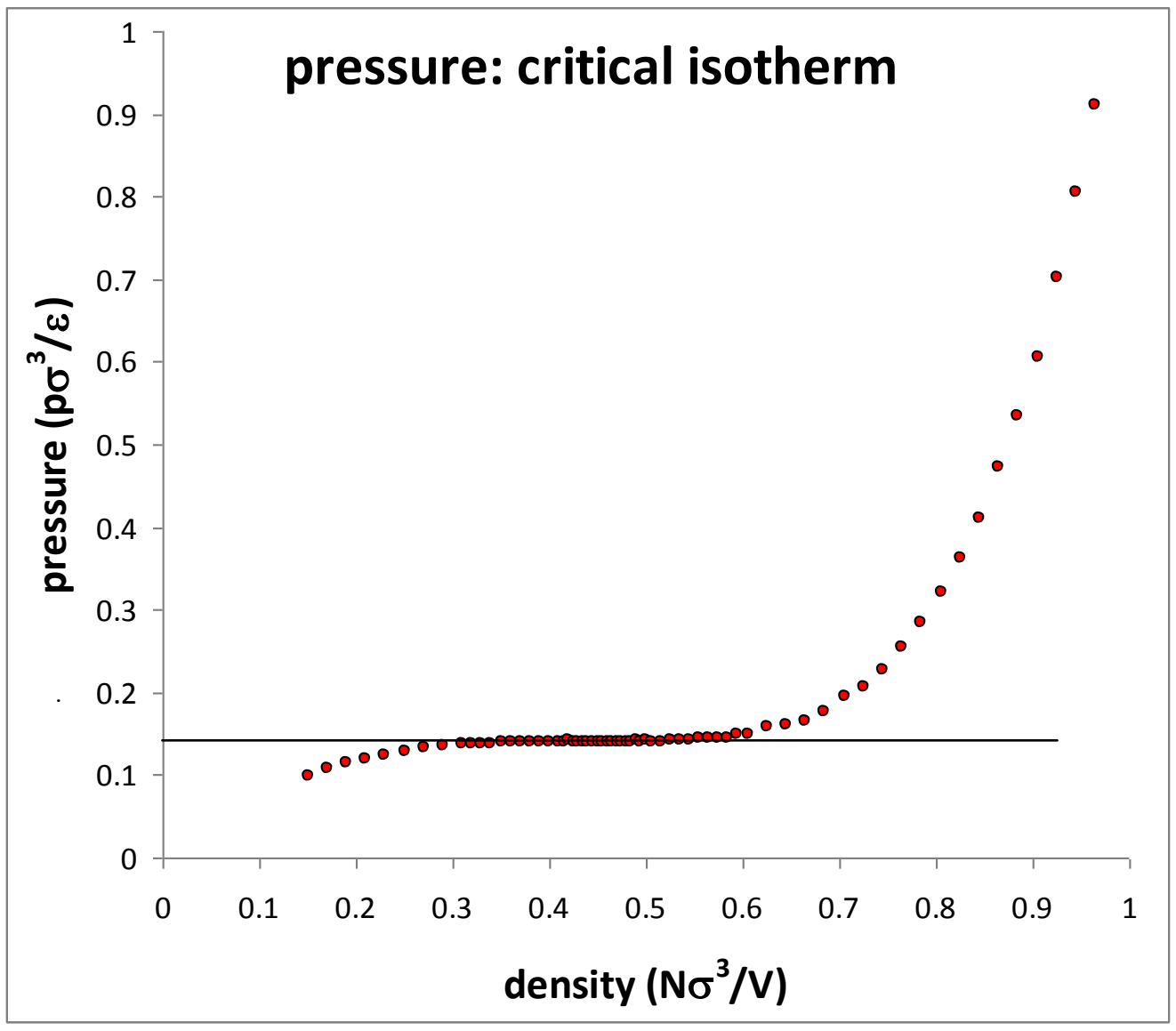

**Figure 4**: MD data averages for the pressure of the Lennard-Jones fluid along the critical isotherm $T^*$ =1.3365 obtained for a system of N= 10976 Lennard-Jones atoms with periodic boundaries.

## CRITICAL COEXISTENCE LINE

The minimum coexisting liquid density and maximum coexisting vapor density at $T_c$ are obtained by interpolation of ($p^*(\rho)$ –$p^*_c$) to zero on either side of the line of critical states. (**Figure 5**). The critical pressure remains constant between two coexisting densities $\rho_c$(gas)$r_o^3$= 0.370±0.01 and $\rho_c$(liquid)$r_o^3$ = 0.527±0.01 which are the densities of the respective percolation transitions PB and PA at the critical temperature.

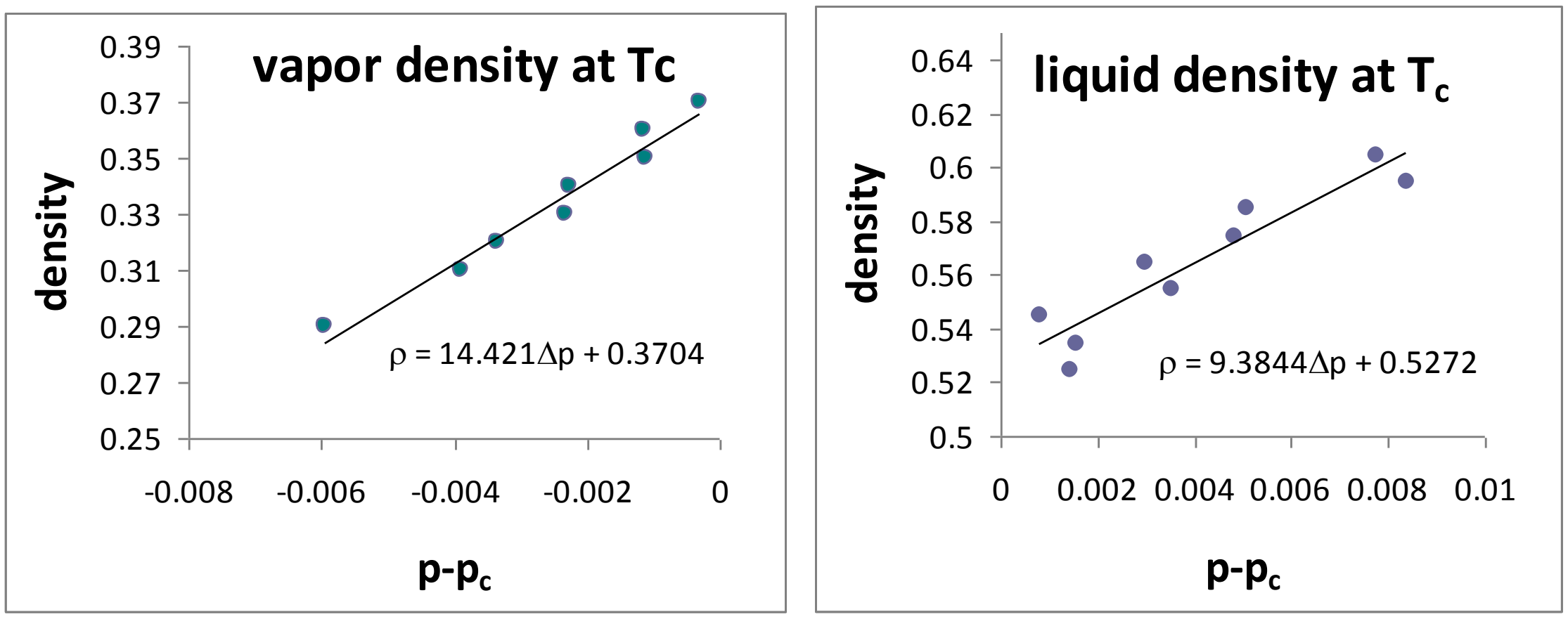

**Figure 5**: Determination of the critical coexisting densities (defined $Nr_o^3/V$) of the Lennard-Jones fluid along the critical isotherm $T_c^*$ =1.3365 using the pressure data in Figure 4 for a system of N= 10976 atoms $\Delta p$= $p-p_c$ where $p_c$ is the critical pressure dimension ($\varepsilon/\sigma^3$)

The total chemical potential calculated by the Widom insertion method [15,16] for the same data points on the critical isotherm is shown in **Figure 6** alongside the critical coexistence densities of liquid and gas. It is constant within a very narrow statistical uncertainty. Thus, in the density range between the PB and PA 2$^{nd}$ order phase transition densities, there is a line of critical states, but no critical point.

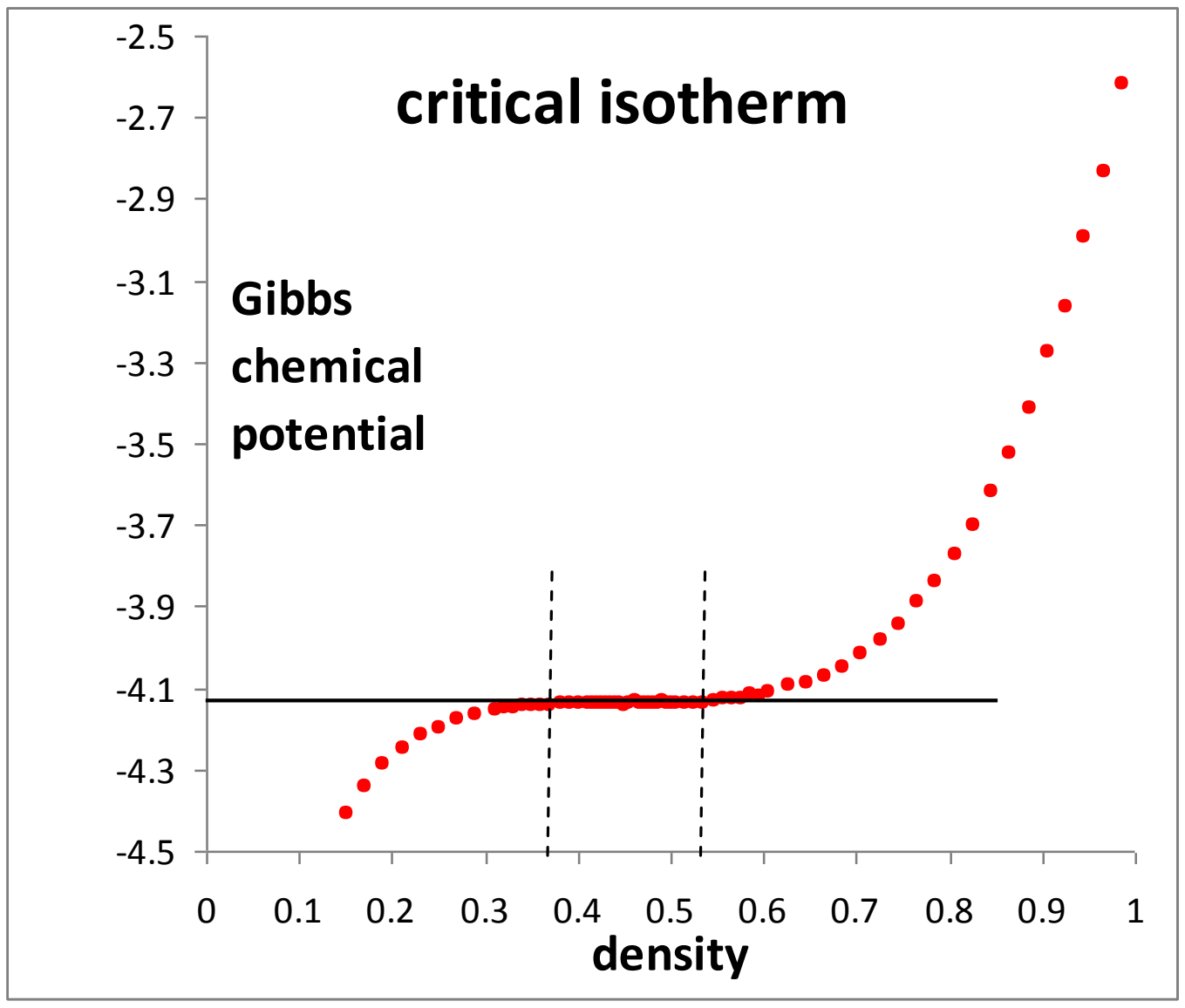

**Figure 6**: Chemical potential of the Lennard-Jones fluid along the critical isotherm $T^*$ =1.3365 obtained using the Widom insertion method for a system of N= 10976 atoms.

Gibbsian thermodynamics strictly only applies to essentially infinite systems in the thermodynamic limit; we can speculate, however, that although in this limit the critical region will be an infinitely thin line, it would not differ significantly from the result that we see here for systems of 4000 and 10976 atoms with periodic boundary conditions.

Above the line of critical states, the supercritical mesophase is homogeneous, with two degrees of freedom, and cannot reduce its Gibbs energy by phase separation into liquid + vapor.

## SUPERCRITICAL MESOPHASE

Using the maximum vapor density obtained above, we can obtain an estimate of a characteristic bond-length that defines a bonded-cluster percolation transition [7,8]. From an EXCEL power-law trendline parameterization of the extended volume percolation transition density as a function of cluster-length λ from table I in the paper of Heyes *et al.* [7], an inversion gives

$$\lambda = 1 + 0.0412\, \rho^{*(-1.617)} \qquad (3)$$

Substituting ρ* for the maximum coexisting vapor reduced density (0.3704) at the critical temperature, we obtain $\lambda_{PB}(T_c) = 1.205$.

MD configurations along the supercritical isotherm $T_c = 1.5$ have been carried out for a system of size N=2048 to determine the probability distribution of clusters in equilibrium configurations. A FORTRAN subroutine CLUSTER for a specifically defined characteristic cluster distance analyzes large numbers of equilibrium configurations for both the average total number of clusters $n_c$, and the probability distribution for a cluster of any size n.

The results for the variation in the total number of clusters with density, $n_c(\rho)$ , reflect the discontinuities at the percolation transitions as can be seen in **Figure 7**. At very low density, i.e. for an ideal gas, there are no clusters and $n_c/N=1$. As the density increases there are an ever-increasing number of pairs, then triplets etc., and increasingly fairly large clusters, until eventually, there is just one cluster of size N that fills the whole box. At this point the probability function $(n_c-1)/N$ goes to zero for large N.

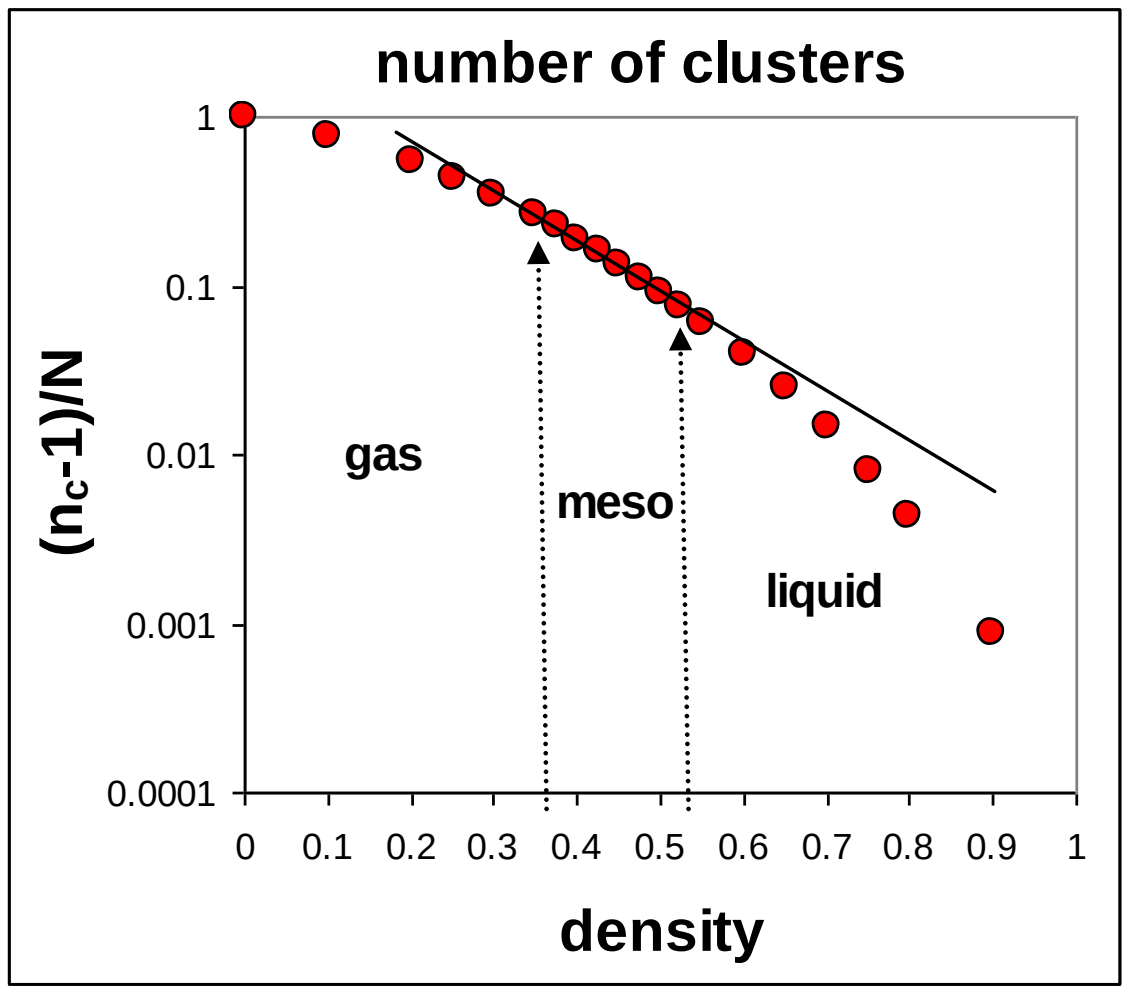

**Figure 7**: Normalized total number of clusters ($n_c$) obtained for the Lennard-Jones fluid at the supercritical isotherm T*=1.5

The function $n_c(\rho)$ obtained from various data points taken from MD studies of the three system sizes investigated, shows a rather weak change in form at the densities of the two percolation transitions. These data give a clue to the percolation densities, but are insufficient to determine it accurately. $n_c(N,\rho)$ has been found to be surprisingly independent of N for the hard-sphere fluid excluded and accessible volume clusters, i.e. for $\lambda=2\sigma$. [8].

More illuminating information regarding the three supercritical subphases can be gleaned from density–dependent % probability function P(n,r)) defined by

$$\%Pr(n) = 100 \text{ x } n_c(n)\ n\ /N$$

Multiplication by 100 is simply to avoid very small numbers on the log plots; the total probability normalizes to 100% i.e. $\Sigma_n P_n=100$. This is a normalized probability of obtaining a cluster of size n. Results for the probability of a cluster of size n are shown in Figure 8 for three densities near the middle of each supercritical region.

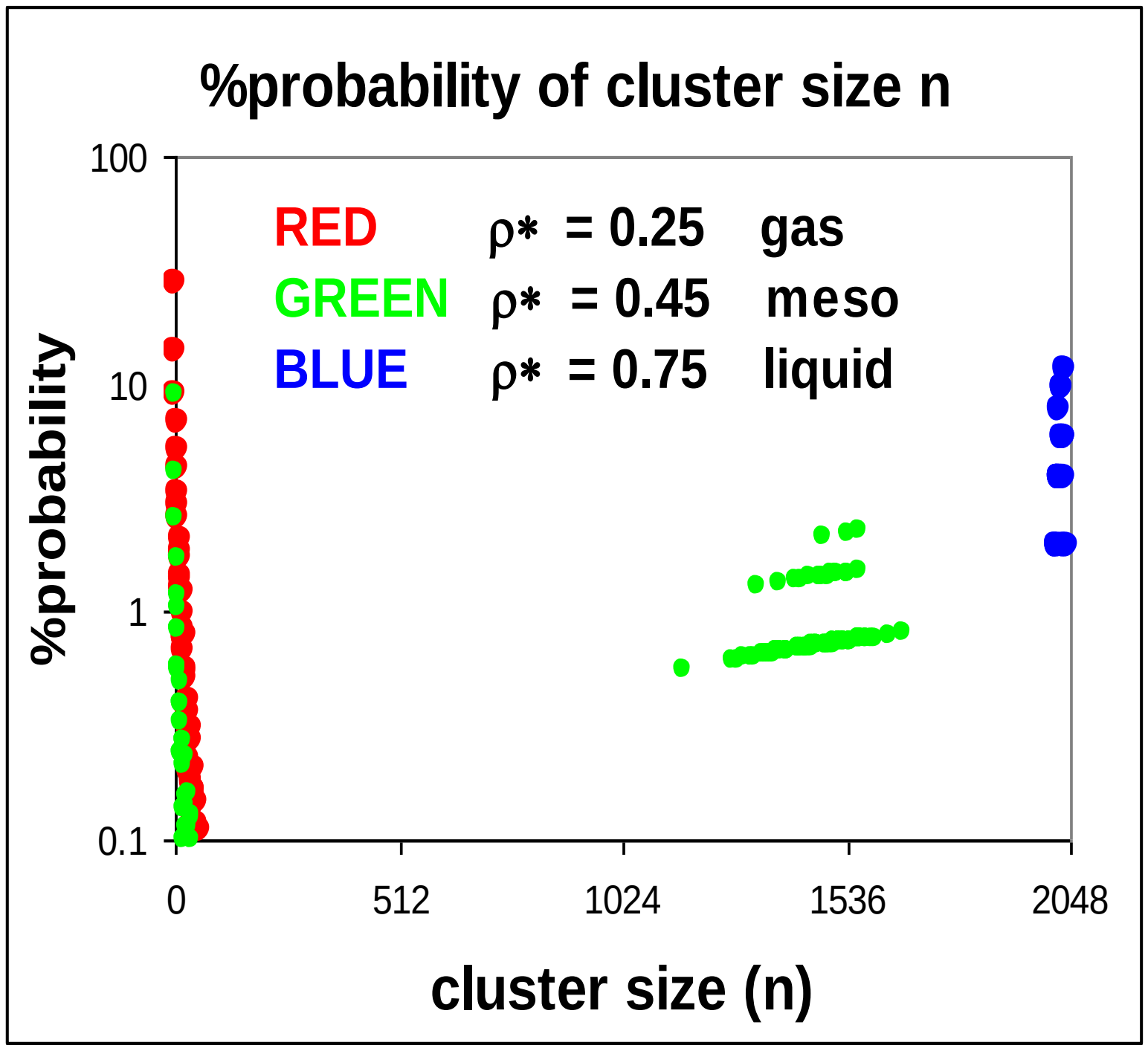

**Figure 8.** Cluster probability frequency for the three supercritical fluid phases

The cluster probability distributions reveal a clear distinction between the three supercritical phases. The gas phase at densities below PB shows a distribution of monomers, dimers, trimers etc, with decreasing probability. The liquid phase, above PB shows just one cluster that is of the order of the size of the system. The mesophase for densities between PB and PA is a combination of both. The data in **Figure 1,** showing constant rigidity $(dp/d\rho)_T$ in the mesophase, suggest a linear combination rule for the respective pressures of the two phases gas and liquid. This is interesting since, if we regard the supercritical meso-state as a multi-component system comprising a mixture of clusters, all cluster species must have the same chemical potential, but the system spanning liquid clusters must have a higher local pressure than the gaseous like species of monomers, dimers etc.

## PERCOLATION LOCI

The implications for our eventual understanding and formal thermodynamic description of the supercritical region are far reaching. Various supercritical lines of extrema of density fluctuations and diffuse discontinuities in second order thermodynamic properties have been reported [17-19]. Brazhkin et al. [20] have calculated the values $C_p$ , $\alpha_p$, and a "Widom line" (related to $\kappa_T$) for several isotherms of the Lennard-Jones supercritical fluid (see also Refs. [21,22]). These authors are unaware of the role of percolation transitions, but an inspection of their results is not inconsistent with the present thermodynamic based description of the percolation loci.

First, $C_p$ against pressure, in Fig. 3a of reference [22], shows a pronounced discontinuity at the reduced pressure $p\sigma^3/\varepsilon = 0.2$ for the lowest isotherm $T/T_c = 1.4$. Second $\alpha_p$ isotherms in figure 2b of Ref. [22] have flat maxima roughly spanning the two percolation densities shown here in **Figures 1 and 3**. The isothermal compressibility ($\kappa_T$) increases monotonically from ideal gas to dense liquid, but when multiplied by a "correlation length", that decreases monotonically from ideal gas to dense liquid, the resulting maximum locus, which has been called the "Widom line", seems to lie broadly between the two percolation densities $\rho_{PA}$ and $\rho_{PB}$, with no particular thermodynamic status. In the supercritical part of the phase diagram there is a low density ('gas') phase to the left of the PB percolation line, and to the right of the PA percolation line there is the liquid phase; which indicates that there can be no single analytic function for the equation of state in the supercritical region of the phase diagram.

Percolation transitions are known to be associated with discontinuities in linear and nonlinear transport properties [8,20,23] and various other dynamical properties such as frequency spectra [24]. There is an increasing literature of hitherto "inexplicable" supercritical lines associated with observations of changes in dynamical properties from gas-like to liquid-like in various supercritical fluids, including water [25-28]. It seems likely a simple thermodynamic explanation for the existence of all these lines will be forthcoming only when the percolation transition loci in these fluids are investigated.

Interestingly, the little graphic presented within the abstract of the paper of Brazhkin et al. [23], shows all three lines of maxima stemming from the critical point in the p-T plane. All of their lines can be identified with the present thermodynamic percolation transition loci, shown here in figure 1 also in the p-T plane. Their $\zeta(T)$-max locus is near PB, their $C_p$-max locus is near PA, and their $\alpha_p$-max locus is intermediate between PA and PB loci. A discontinuity locus referred to by Brazhkin et al. [20] as a "Widom line" can actually be seen in Figure 9 of the paper by Heyes and Melrose published 25 years ago, which was identified as the line a percolation transition [6].

## DISCUSSION AND CONCLUSION

The van der Waals theory of a critical point [1], for which he received the Nobel prize in 1910 [29], has been universally accepted description of liquid-gas criticality ever since. According to this hypothesis, every fluid has an equation-of-state for which the first two derivatives of pressure with respect to changes in density or volume, go to zero at a singular point on the Gibbs density surface. A second misapprehension that has been in fashion for 50 years, a concept to describe all critical phenomena, from Ising models, ferro-magnetic systems, spin glasses add liquid-gas criticality, is 'universality'. All critical phenomena, according to this theoretical physics community, obey the same universal physical description, culminating in a second Nobel prize being awarded in 1982 to Wilson [30] for his application of renormalization group theory to the generalization of critical points descriptions including the hypothetical van der Waals 'critical point' singularity.

In the Nobel addresses: van der Waals had problems with critical volumes, he refers throughout his Nobel lecture to "the critical volume", but nowhere does he say that its existence is a hypothesis; i.e. it is very misleading. At one stage, however, he states: "…..and again it can be seen from the equation that what I have termed the weak point of my theory is actually responsible for the theoretical impossibility of calculating accurately the critical volume". In fact, it is the incorrect hypothesized mathematical form of the equation-of-state of supercritical fluids, with its cubic node, that generates the spurious 'critical point'.

Wilson, in the introduction to his Nobel address [30], summarizes the phenomenology of liquid-gas criticality by reference to water and steam in one brief paragraph as follows: "A critical point is a special example of a phase transition. Consider, e.g. the water-steam transition. Suppose the water and steam are placed under pressure, always at the boiling temperature. At the critical point, the distinction between water and steam disappears, and the whole boiling phenomena vanishes. The principal distinction between water and steam is that they have different densities. As the pressure and temperature approach their critical values the difference in density between water and steam goes to zero".

This statement is the only reference to any liquid-gas experimental coexistence data in Wilson's entire lecture. There is no reference at all to the Gibbs classical thermodynamic description, in a 32-page Nobel lecture. In fact, liquid-gas criticality is not mentioned again! We now see that the general experimental phenomenology on which Wilson's theory, and indeed also the Nobel prize, is to a large extent, justified, is quite incorrect.

Notwithstanding two Nobel prizes in physics for theory of a 'critical point' that does not exist, we conjecture that the present description of liquid-gas criticality of the Lennard-Jones fluid will extend to all thermodynamic fluids, defined by Gibbs thermodynamic limit and as being capable of both reversible heat and p-V work. The density difference between water and steam is the same phenomenology [4], it does not go to zero at the critical temperature. Moreover, for temperatures above the critical temperature the gas and liquid phases are separated by a supercritical mesophase which is a homogeneous mixture of gas-like small clusters, and system spanning large clusters, all with the same Gibbs chemical potential. These phase boundaries appear as a weak second-order thermodynamic phase transition, referred to as percolation transition loci, not presently well understood at the molecular level, but now requiring further experimental and theoretical investigation.

Finally, as a preliminary observation, we have noted that the supercritical percolation lines are consistent with the continuation of the vapor and liquid metastable limit spinodal lines into the supercritical region to bound the density surface existences of gas and liquid phases, rather than ending, or merging, at a hypothetical 'critical point'. Thus, all aspects of van der Waals proposed function for the equation-of-state, not just the cubic node singularity, appear to be incorrect in the vicinity of $T_c$. Indeed, we now see that, above $T_c$, there will be three different analytic functional forms for the equation-of-state of a supercritical fluid. Both gas and liquid equations-of-state should extend down to triple point temperatures, and indeed beyond for the gas and supercooled liquids, and also into the metastable existence regions up to the spinodal instabilities.